\documentclass{PoS}

\title{Highlights of EPS HEP 2019}

\ShortTitle{Highlights}

\author{\speaker{Jon Butterworth}\\
        Department of Physics \& Astronomy, University College London\\
        E-mail: \email{j.butterworth@ucl.ac.uk}}

\abstract{An opinionated and informal recap of highlights from the EPS HEP 2019 conference
in Ghent, including some aspects of flavour physics, neutrinos, high-density QCD, astrophysics and energy frontier 
collider physics, and some thoughts about the future.}

\FullConference{
European Physical Society Conference on High Energy Physics - EPS-HEP2019 -\\
			10-17 July, 2019\\
			Ghent, Belgium}

\begin{document}

\section{Introduction}

The purpose of this write-up, as with the talk, is not to be definitive but to give an editorialised point 
of view on what was a packed and exciting meeting. As befits such a contribution, references will mostly be to other (mostly plenary)
talks at the conference, and I will not include figures, on the assumption these will be included in the original contributions. 
Although it is a necessarily 
personal selection, I will not dwell on a satisfying semi-final and a stunning final from the cricket world cup, 
which took place during the conference, even though they will always be associated with Ghent for me from now on. 
I will say though that it was a privilege to be asked to give this talk by
the European Physical Society, especially bearing in mind the turbulent politics between the UK and most of the rest of Europe at 
present.

For the rest of the  contribution, I will stick to the physics. There was a lot of it.

\section{Flavour Physics}

Impressive advances were reported in flavour physics, with Katharina M\"uller presenting the first observation of charge-parity
violation in charm decays, by LHCb~\cite{Aaij:2019kcg}. The time-integrated CP asymmetry is measured in neutral charmed hadron ($D^0/\bar{D}^0$)
decays to both $\pi^+\pi^-$ and $K^+K^-$. A combination of the new LHC Run 2 result with the Run 1 result gives a 5.3$\sigma$ difference of the 
measured value from zero, indicating dominantly direct CP violation and roughly compatible with the Standard Model (SM). 
The uncertainties in the SM prediction are greater than those in the data.

M\"uller also presented an updated combination of measurements of the unitarity triangle angle $\gamma$, incuding a new measurement
in the $B^0 \rightarrow DK^{*0} (D \rightarrow K\pi, KK, \pi\pi)~$\cite{Aaij:2019tmn}. There are some tensions, at the 2$\sigma$ level, 
between different measurments, worth mentioning to emphasise the importance of measuring this angle in several different ways, since
new physics may manifest itself differently depending upon the process. Updates of $B_s$ mixing from LHCb and ATLAS were also presented, using
integrated luminosities of 4.9$^{-1}$ and 99.7~fb$^{-1}$ respectively. These results were also discussed by Johannes Albrecht. 
The main focus in his talk however was on semi-leptonic decays of $B$ hadrons to Kaons and either $e^+e^-$ or $\mu^+\mu^-$, 
and in particular the 
ratios and angular variables, where the consistency with the SM is at level of around 2.5$\sigma$, a level which has not changed 
with the increased precision from including part of the Run 2 data. While no individual measurement is compelling as evidence for physics
beyond the SM (BSM), taken together (and as pointed out by Marco Nardecchia in his summary of these and other comparisons) 
there is a suggestive pattern which is readily accommodated by extensions to the SM. This is clearly an interesting space to watch, especially
as more LHC luminosity is included and, as we heard from Francesco Forti, Belle II data are on the horizon, with the accelerator 
currently on its commissioning track to high luminosities and the detector already recording collisions.

\section{Neutrinos}

That we are now into an era of precision, high statistics neutrino physics was amply illustrated by the dataset 
of atmospheric neutrino events collected by SuperKamiodande~\cite{Abe:2017aap} and now used in global fits, with 
a ``postage stamp'' style slide shown by Gioacchino Ranucci containing
50,000 events split into 19 different analysis samples, each with enough statistics to allow detailed exploration of the event
characteristics. As discussed by Francesca Di Lodovico and Sylvia Pasquale, data principally from T2K and No$\nu$a are beginning 
to constrain both the neutrino mass hierarchy (with the normal
hierarchy marginaly favoured) and the CP violating phase in the PMNS matrix (with a non-zero value marginally favoured).
Reactor experiments are also making major contributions to characterising the physics of the neutrino sector, particularly 
with their measurements of the mixing angle $\theta_{13}$.

\section{New ``Standard Model'' Physics}

When we discuss the search for ``new physics'', it grates on me sometimes, since as we probe above the electroweak symmetry breaking scale
with ATLAS and CMS, everything we measure is in a sense new. We are measuring what are sometimes previously unseen processes, and always
in a region of physics previously unexplored. Every time we do so and find agreement with SM, we are validating our ideas about the
fundamental constituents and forces in a qualitatively 
new regime. By the standards of most fields, this is new physics.

Several examples were shown by Wolfgang Adam, Lucia Di Ciaccio and Andreas Hoecker, with (to give just a few examples) 
impressive results on diboson production, top cross sections and asymmetry (including new calculations of NNLO spin correlations),
prompt photons (now out to 2~TeV in transverse momentum) and a dilepton+photon measurements with the full Run 2 data set from ATLAS.
As Giulia Zanderighi made clear, for the full exploitation of these measurements our ability to make exclusive calculations, 
implementing the kinematic cuts which allow comparisons to data in the fiducial phase space of the measurement, is essential and is 
an area of intense effort and much progress.

The new regimes probed, and the new precision achieved with QCD theory, means that, as also pointed out by Zanderighi, 
there is now a need for both QCD and electroweak higher order corrections to be incorporated together in calculations. This
need will become more urgent throughout the high luminosity LHC (HL-LHC) era. An early step in this direction was in fact shown in 
a parallel session by German Sborlini, in the form of a coherent QED and QCD higher-order calculation of Drell-Yan dilepton 
production~\cite{Cieri:2018sfk}. It seems from this that a multiplicative rather than additive combination of the corrections 
is a better approximation to the fully coherent result.

At the time, I thought I understood something about elliptical Feynman diagrams from the fascinating talk by Claude Duhr. 
This understanding now escapes me, but it seems there is an interesting and useful advance on the boundaries between physics 
and mathematics there.

As predicted, jet substructure analysis has proved itself {\it ``useful more generally in the identification of hadronically decaying massive
particles which have energy large compared to their mass''}~\cite{Butterworth:2002tt}, and highly boosted $W,Z, H$ and top are 
now an important feature of physics at the LHC. Several new results were shown, including a very direct technique from CMS 
of measuring the top mass by measuring the mass of a ``fat'' jet containing three subjets from the decay of a top quark.
There are interesting questions about how these techniques will change, and what demands they will place on the detectors,
if even higher boosts are available at future colliders, and when also the scales are such that these particles will need to be treated as 
partons in the initial and final state showers. Present theoretical tools are likely to require substantial
development for us even to make a reasonable estimate.

\section{High Density QCD}

Jet substructure also featured as a powerful tool in heavy ion physics and the study of high-density QCD. An array of impressive 
new results was shown by Marco Van Leeuwen, Dennis Perepelitsa and Carlos Albert Salgado Lopez. The impact of the medium on QCD
splitting is studied by observing differences between the subjet momenta in proton-proton and heavy ion 
collisions~\cite{Sirunyan:2017bsd,Acharya:2019djg}, with interesting effects seen which should help characterise the medium. However,
such comparisions must be done with care, given the obersvation of collective effects even in proton-proton 
collisions~\cite{Khachatryan:2010gv}, a system thought, 
before LHC data, to be too small to manifest such a thing. As Jan Fiete Grosse-Oetringhaus strikingly put it, this challenges 
two paradigms at once. What is the smallest system in which the heavy-ion “standard model” remains valid?
And can the standard tools for proton-proton physics remain standard? Traditional high-energy physics and traditional heavy ion studies,
often sociologically and scientifically rather distant, must grow together. The underlying QCD is the same theory, and the aim must
be to demonstrate that unified description for $ee$, $ep$, $pp$ and $AA$ is feasible, or show that different mechanisms are justified
and how they arise.

\section{The Higgs}

Even if you don't buy my claim that most of the measurements from the LHC are new physics, perhaps you will agree
with Roberto Salerno's statement that Higgs physics, at least, is {\it really} new physics. Christoph Grojean in
the ECFA session on Sunday gave a powerful reminder, in the context of proposals for future colliders~\cite{deBlas:2019rxi}, 
of just how special the Higgs boson is. It can be seen as a new force, of a different nature to the gauge interactions 
known so far. There is no underlying local symmetry, no quantised charge. It is deeply connected to the vacuum structure 
of space-time. The up and down-quark Yukawa couplings determine the relationship between the proton and neutron masses, and 
thus the stability of nuclei. The electron Yukawa controls the size of the atoms (and thus the size of the Universe?).
The top quark Yukawa decides (in part) the stability of the electroweak vacuum. The Higgs self-coupling controls the 
(thermo)dynamics of the EW phase transition, and therefore might be responsible for the dominance of matter over antimatter in the
Universe. The Higgs Boson really is special. 

Salerno, Adam and Hoecker showed a plethora of new results, many using the full LHC Run 2 dataset. ATLAS and CMS have established 
the existence of Higgs couplings to the third generation 
charged fermions, as well as the gauge bosons. New studies were shown on the pursuit of the second generation --- muons and charm. The
limits are still factors above the SM, but are getting close enough to have interesting sensitivity to any upward deviations, and 
more precision is promised. Differential cross sections for Higgs production and decay test whether is it pointlike, whether its 
couplings evolve as expected with (for example) transverse momentum. Direct measurement of the Higgs self coupling requires much more data, 
with observation possible with the full HL-LHC, and any precision requiring a future hadron collider, just as a measurement of the 
total Higgs width with a degree of model-independence requires a lepton collider. 

Finally on the Higgs, Francesco Riva detailed how intertwined the Higgs sector is with many of the processes measured at the LHC -- dibosons,
top and more. This implies that when we measure such things, we are, within the context of the SM, making powerful consistency tests
of the parameters of the Higgs sector -- or doing Higgs physics without a Higgs, as he styles it. So, if you agreed that Higgs physics
is new physics, you now have to accept my earlier claim that so are many of the other measurements ATLAS and CMS are producing, even if
no trace of BSM physics has yet shown itself.

\section{Dark Matter, Astroparticle Physics and Cosmology}

The progress in searching for Dark Matter was described by Igor Garcia Irastorza, Carlos de los Heros and Kfir Blum.
Amongst the items which were fresh news to me was the fact that the famous picture of the black hole at the centre of 
the giant elliptical galaxy M87~\cite{Akiyama:2019cqa} actually excludes some ultra-light Dark Matter models, 
in which such black holes do not form. This is partly I think a 
reflection of a resurgence in model building and exploration, now the WIMP miracle is (at best) delayed. This is also
reflected in the increased interest in axion searches as described by Irastorza.

We also learned that the so-called ``neutrino floor'', the point in sensitivity at which detectors looking directly 
for Dark Matter scattering off normal matter start seeing solar neutrino interactions, is not a hard floor, ``more of a swampland'' in 
the words of de los Heros. Anisotropic detectors, directional detection and time modulation offering possibilities for sinking
below it. 

The anomalously large number of positrons seen by AMS (shown by Barbara De Lotto) remains intriguing -- something interesting is 
going on there, whether it is Dark Matter-related or not. And the IceCube map of the neutrino sky shown by Elisa Bernardini 
offers a truly new view of the Universe, with the obvious potential for identifying point sources perhaps from Dark Matter annihilation, 
amongst other things.

Our newest messenger, gravitational waves, continues to surprise and excite.
One surprise to me was the fact that the ``chirp'' pattern of a neutron star merger, as shown by Patricia Schmidt, has 
-- via tidal distortion -- implications for the equations of state of high density QCD, as discussed by Carlos Albert Salgado Lopez.
The prospect of the Einstein telescope, which with its sensitivity to lower frequencies should allow impending 
mergers to be spotted earlier and thus observed in more channels, is mouth-watering.

\section{The Future, and Beyond the Standard Model}

And so to something on future prospects. Much of this hangs on development of accelerator technology, of which we heard 
summaries from Catarina Biscari in the ECFA session and Ralph Assman in the final session of the plenaries.
Highlights included operational crab cavities for protons in the SPS (shown by Olivier Bruning in the parallel sessions), 2 GeV electron 
acceleration in the AWAKE proton-driven plasma wakefield experiment, a dipole magnet demonstrator so far tested at FNAL to 14 Tesla with
an end goal of 15 T, and work on Nb$_3$Sn wire toward 16 T from companies engaged in the Future Circular Collider project. The success (and
speed) of such developments will be critical for the long-term future of the field. 

Nearer in time, we have LHC Run 3 and the HL-LHC era approaching. 
Isabell Melzer-Pellmann and Marie-Helene Genest showed the results of many ingenious and important searches for BSM physics, none
of which have revealed anything other than increasingly stringent limits. This is itself is great progress and is having a profound impact
on the theoretical field, as was described by Giuliano Panico and alluded to several times already in this contribution.
Over the next few years, I believe we need, and will see, a profound change of approach here.

In my opinion, while there is still important scope for novel signatures not yet covered (for example, exotic 
long-lived particles, disappearing tracks, non-standard jets, and probably other things still to be dreamt up) 
the main emphasis of the experiments should shift toward making precise, model-independent
measurements which can be confronted with increasingly precise and exclusive SM predictions. It
is such confrontation in my view that gives us the best chance of establishing whether or not we face a 
desert above the electroweak symmetry-breaking scale, as far as BSM physics is concerned. 
Such measurements, including those of the Higgs, seem likely to me to provide the main legacy of the LHC, 
an exacting challenge laid down to future model builders.

A clearer idea of exactly what is being measured, in terms of separating theoretical interpretation from measurement, is needed. 
This implies not only measuring in fiducial regions reflecting the detector acceptance, but also in what is considered background 
and what is considered signal. An example: if we are aiming to measure $WW$ scattering, and we see an event with two isolated leptons and 
missing energy in the detector, 
should we really be subtracting it because our Monte Carlo says it came from top quarks? What about off-shell tops? It is often better to 
measure a final state, and do any subtraction later as part of the interpretation.
This throws down a challenge to the theory to calculate the final states that are actually measured, of course, as discussed above.

When it comes to interpretation of searches too, we need to on the one hand provide the data in such a way that a search for one
model may be readily reinterpreted in terms of others with similar signatures, and on the other hand we should be moving away
from simplified benchmark models toward making more general statements about whole classes of theory. An excellent example of the latter
approach was given by Peter Athron in his parallel session talk~\cite{Athron:2019fdk}, where the GAMBIT system is used to show that
rumours of the death of the MSSM may be exaggerated. Not only are there still allowed parameter points for any and all values of neutralino 
and chargino mass (even though swathes of the multidimensional parameter space are indeed ruled out by LHC data) but there are even some 
regions of SUSY parameter space which are marginally favoured over the SM. 

I will take the opportunity to highlight a few other techniques and ideas of growing importance.

The potential of machine learning, long extant around the periphery of the field, is now being realised in many different 
particle and astroparticle physics applications, and has much more to offer if used judiciously.
Jet substructure has already been mentioned and hardly counts as new, but will in my view continue to grow in importance, including
in the evaluation of the potential of future colliders. The same goes for the theoretical advances needed to calculate such variables
incorporating both QCD and electroweak higher-order corrections.

And finally, the field of particle physics really must get to grips with Open Data. The culture in astrophysics should be an example; the 
fact that LIGO data from the first gravitational wave observations is already public is immensely impressive, when the Higgs discovery data 
are still behind collaboration firewalls. Of course, the technical issues are different. Release of
data is not without effort, and brings challenges. But in the final analysis, data produced by our experiments is part of the 
store of human knowledge, and is not ours to keep to ourselves in the long term. On the LHC side, 
CMS have led the way in opening a subset of early 
LHC data for physics analysis, some of which has been used by theorists to develop new results, and as shown in a parallel session
by Matthias Schott, a member of ATLAS, to perform open cross-checks, new measurements and comparisons. This shows opening up our data is
technically possible, and useful, for collider experiments. However, such activity -- especially by a 
member of a rival collaboration -- causes concern to some, and may even be discouraged or seen as a threat by the collaborations.
This would be a mistake. 
While there are, for example, legitimate concerns about shortage of effort within the collaborations, 
attempting to control what colleagues do with public
data is not the way to address those concerns, and in my opinion crosses an unacceptable line in terms of academic freedom.
Open data, and its analysis, will in the end be good for science and innovation, and should be at least tolerated by even the most 
sceptical. External pressures are pushing the field in this direction anyway. We should go enthusiastically, not dragging our feet.

\section{Conclusion}

With the ongoing results from the LHC experiments, Belle II arriving, HyperK and DUNE on the horizon,
continuing improvement in theoretical and experiment techniques and a deepening engagement between particle physics,
astrophysics and nuclear physics, this week has provided an exciting snapshot of a field still digesting the implications of 
the results of the last decade or so. I did not mention charged-lepton flavour measurements, or experiments probing the neutrino 
mass directly (its value, and whether it is Majorana or Dirac in nature), but there are also exciting prospects there.
The main impression I took away from the conference is that while we need to pay attention
to the far future, and it contains many uncertainties, the next few years promise a great richness of
data, and therefore possibility, which should inspire and excite us all. But there is still much work 
to do to make this a reality.

I apologise to anyone I didn't represent well, or at all, here. 
I already thanked the organisers for the invitation. I would also like to thank them and the local organisers, 
colleagues, and the city of Ghent and its brewers, for hosting and attending a superb conference.

\end{document}